\documentclass[preprint,5p,times,twocolumn,nopreprintline]{elsarticle}
\usepackage{amsmath,amssymb,amsfonts}
\usepackage{graphicx,color,float}
\usepackage{subfigure}
\usepackage{todo}
\usepackage{tikz}
\usepackage{multirow,booktabs}
\usepackage{epsfig}
\usepackage{graphicx}
\usepackage{slashed}
\usepackage{bm}
\usepackage{amsmath} 
\usepackage{amssymb}
\allowdisplaybreaks

\newcommand{\half}{\mbox{${\textstyle \frac{1}{2}}$}}           
\newcommand{\fourth}{\mbox{${\textstyle \frac{1}{4}}$}}         

\newcommand{\rd}{\textrm{d}}
\begin{document}
\begin{frontmatter}
\title{Hadronic structure functions in the $e^+ e^- \rightarrow   \bar{\Lambda} \Lambda$ reaction}
\author[IKPUU]{G\"oran F\"aldt}
\ead{ goran.faldt@physics.uu.se} 

\author[IKPUU]{Andrzej Kupsc}
\ead{andrzej.kupsc@physics.uu.se}

\address[IKPUU]{Division of Nuclear Physics, Department of Physics and 
 Astronomy, Uppsala University, Box 516, 75120 Uppsala, Sweden}

%
%
  
%
%
\date{Received: \today / Revised version: date}

\begin{abstract}
Cross-section distributions are calculated for the reaction 
$e^+ e^- \rightarrow  J/\psi\rightarrow  \bar{\Lambda}(\rightarrow \bar{p}\pi^+) \Lambda(\rightarrow p\pi^-)$,
and  related annihilation reactions mediated by vector mesons. The  hyperon-decay distributions 
depend on a number of structure functions that are bilinear in the, possibly complex, 
psionic form factors  $G_M^\psi$  and $G_E^\psi$ of the Lambda hyperon. 
The relative size and relative phase of these form factors can be uniquely  determined from the 
unpolarized joint-decay distributions of the Lambda and anti-Lambda hyperons. 
Also the decay-asymmetry parameters of Lambda and anti-Lambda hyperons can be determined.

\end{abstract}


\end{frontmatter}

%
%
%
\section{Introduction}\label{ett}

\begin{figure}[ht]
a)\hfill

\centerline{\scalebox{0.60}{ \includegraphics{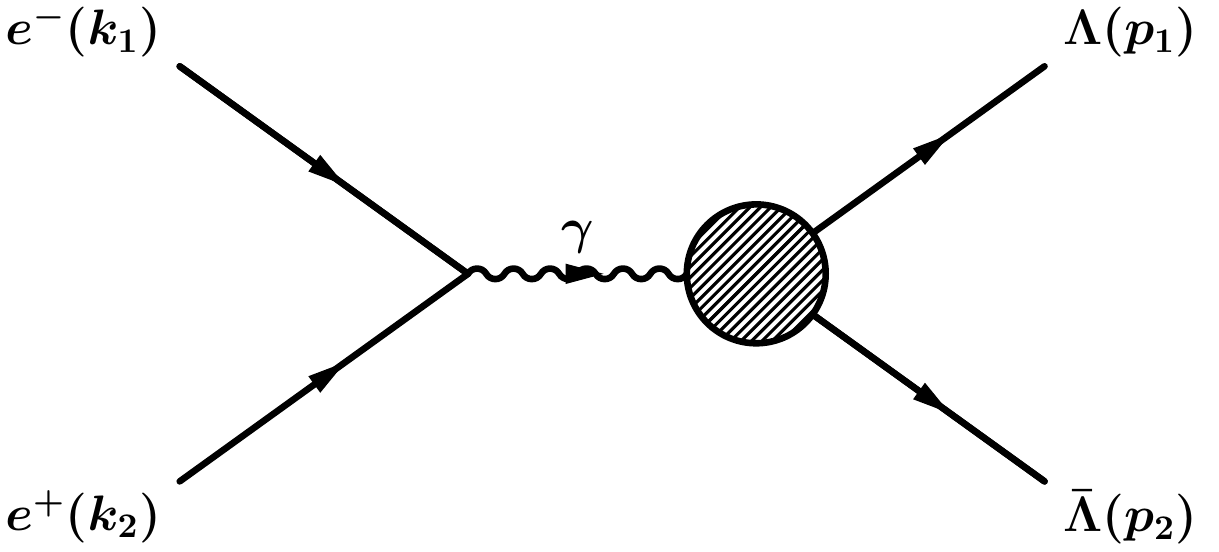} }}    

b)\hfill

\centerline{\scalebox{0.60}{ \includegraphics{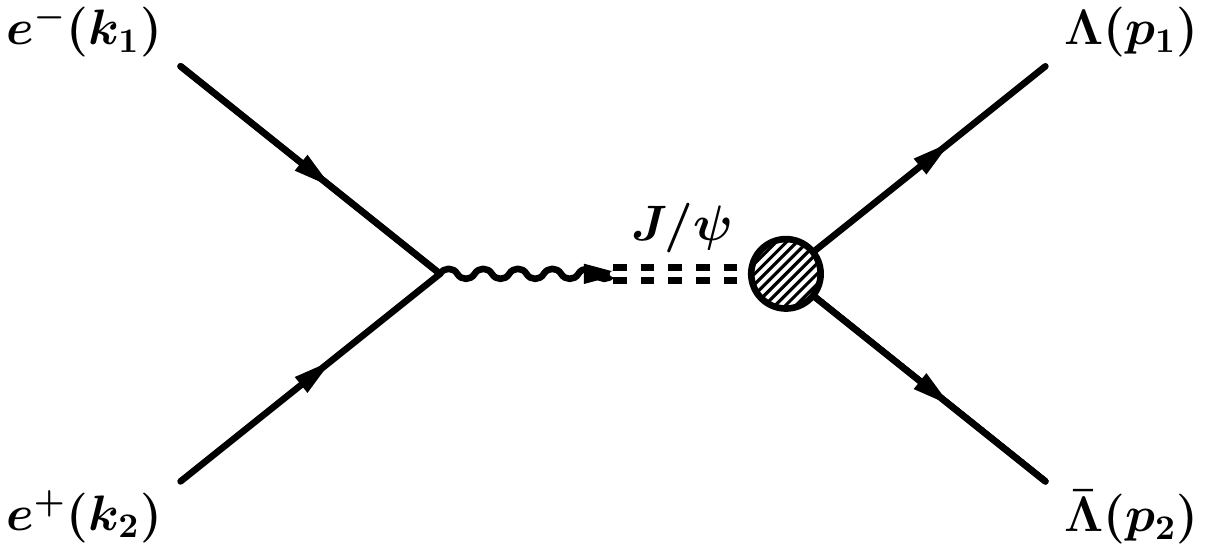} }}    

\caption{Graph describing the reaction 
$e^+ e^- \rightarrow  \bar{\Lambda} \Lambda$; a) genaral case, and b) mediated by the $J/\psi$ resonance.}
\label{F1-fig}
\end{figure}

Two hadronic form factors, commonly called $G_M(s)$  and $G_E(s)$, are needed for
the description of the annihilation process $e^- e^+ \rightarrow  \Lambda \bar{\Lambda}$, Fig.~\ref{F1-fig}a, 
and by varying the c.m.\ energy $\sqrt{s}$, their numerical values can in principle be determined for 
all $s$ values above $\Lambda \bar{\Lambda}$ threshold. 
For the general case of annihilation via an intermediate photon,
the joint $\Lambda(\rightarrow p\pi^-) \bar{\Lambda}(\rightarrow \bar{p}\pi^+)$ 
decay distributions were calculated and analyzed in Ref.\cite{GF2},  using methods developed in
\cite{GF1,Czyz}. 
Recently, a first attempt
to calculate the hyperon form factors $G_M(s)$ and $G_E(s)$  in the time-like region was 
reported in Ref.~\cite{Haid}. 

Previously, the interesting special case of annihilation through an 
intermediate $J/\psi$ or $\psi(2S)$, Fig.~\ref{F1-fig}b,
has been investigated in several theoretical \cite{HC-RG,BinZ} and 
experimental papers \cite{DM2a,DM2b,BESa}. 
This process has also been used for 
determination of the anti-Lambda decay-asymmetry parameter and for 
CP symmetry tests in the hyperon system.
A precise knowledge  of the Lambda decay-asymmetry parameter is
needed for studies  of  spin polarization in $\Omega^-$,
$\Xi^-$, and $\Lambda_c^+$ decays.

Presently, a collected data sample of $1.31 \times 10^9$ $J/\psi$ events
\cite{Ablikim17a} by the BESIII
detector~\cite{Ablikim:2009aa} permits high-precision studies of
spin correlations.

In the experimental work referred to above, 
the joint-hyperon-decay distributions considered are not  the
most general ones possible, but  seem to be curtailed. Incomplete distribution functions 
do not permit a reliable determination of the form factors and we therefore suggest
to fit the experimental data to the general distribution described in \cite{GF2},
and further elaborated below.

Since the photon and the $J/\psi$ are both vector particles, their corresponding
annihilation processes will be similar. 
In fact, by a simple substitution,  the cross-section distributions in 
Ref.~\cite{GF2}, valid in the photon
case, are transformed into distributions valid in the $J/\psi$ case, but
expressed in the corresponding psionic form factors  $G_M^\psi$  and $G_E^\psi$.

In order to  specify events and compare measured data with theoretical 
predictions, we need distribution functions expressed in some specific coordinate system.
For this purpose we employ the coordinate system introduced in \cite{GF2}. 
Many investigations employ different 
coordinate systems for the Lambda and anti-Lambda decays, a custom which in our opinion can
lead to confusion.

Our calculation is performed in two steps. After some preliminaries we turn to 
the inclusive process of
lepton annihilation into polarized hyperons. The results obtained are  
the starting point
for the calculation of exclusive annihilation, i.e.~the distribution for the hyperon-decay products.
Our method of calculation consists in multiplying the hyperon-production distribution 
with the hyperon-decay distributions, averaging over intermediate hyperon-spin directions.
The method is referred to as folding.
%
%
%
%
\section{Basic necessities}\label{två}

Resolving the hyperon vertex in Fig.1a uncovers a number of contributions. The one of interest 
to us is described by the diagram of Fig.1b, whereby the photon interaction with the hyperons 
is mediated by the $J/\psi$ vector meson, and the coupling of the
initial-state leptons to the $J/\psi$   related to the decay $J/\psi\rightarrow e^+e^-$. 

For a $J/\psi$ decay  through an intermediate  photon, tensor
couplings can be ignored. Thus, the effective coupling of the $J/\psi$ to the leptons is 
the same as that for the photon, provided we replace the electric charge $e_{em}$ by a coupling 
strength $e_\psi$,
\begin{equation}
	\Gamma_\mu^e(k_1,k_2)=-ie_\psi \gamma_\mu,
\end{equation}
with $e_\psi$ determined by the  $J/\psi\rightarrow e^+e^-$ decay (see Appendix).

At the $J/\psi$-hyperon vertex two form factors are possible and they are both considered.
 We follow  Ref.~\cite{GF2} in writing 
the hyperon vertex  as
\begin{equation}
		\Gamma^\Lambda_\mu(p_1,p_2) = -i e_g\Big[
		G_M^\psi\gamma_\mu 
		  -\frac{2M}{Q^2}(G_M^\psi-G_E^\psi ) Q_\mu \Big], \label{Lambdavertex}
\end{equation}
with $P=p_1+p_2$, and $Q=p_1-p_2$, and $M$ the Lambda mass.
The argument of the form factors equals  $s=P^2$.
The coupling strength $e_g$ in Eq.(\ref{Lambdavertex}) is determined by the  hadronic-decay rate 
for  $J/\psi\rightarrow \Lambda\bar{\Lambda}$ (see Appendix).

In Ref.~\cite{GF2}  polarizations and cross-section distributions were expressed in terms
of structure functions, themselves  functions of the form factors $G_M^\psi$ and $G_E^\psi$.
Here,  we shall introduce combinations of form factors called 
$D$, $\alpha$, and $\Delta\Phi$, 
which  are employed by the experimental groups \cite{DM2a,DM2b,BESa} as well.

The strength of form factors is measured by $D(s)$,
\begin{equation}
	D(s)=s\left| G_M^\psi \right|^2 + 4 M^2 \left| G_E^\psi \right|^2. \label{DS_def}
\end{equation}
a factor that multiplies all cross-section distributions. The ratio of  form factors
is measured by $\alpha$,
\begin{equation}
	\alpha= \frac{s\left| G_M^\psi \right|^2 - 4 M^2\left| G_E^\psi \right| ^2}
	{s\left| G_M^\psi\right|^2 + 4 M^2\left| G_E^\psi\right|^2},\label{alfa_def}
\end{equation}
with $\alpha$ satisfying $-1\leq \alpha \leq 1$.
The relative phase of form factors is measured by $\Delta\Phi$, 
\begin{equation}
	\frac{G_E^\psi}{G_M^\psi}=e^{i\Delta\Phi} \left| \frac{G_E^\psi}{G_M^\psi}\right|. \label{DPHI_def}
\end{equation}

 The diagram of Fig.~1 represents a  $J/\psi$ exchange of momentum $P$. 
$J/\psi$ being a vector meson, its propagator takes the form
\begin{equation}
	\frac{g_{\mu\nu}- P_\mu P_\nu/m_\psi^2}{s-m_\psi^2+im_\psi\Gamma(\psi)},
\end{equation}
 where $m_\psi$ is  the  $J/\psi$ mass, and $\Gamma(\psi)$ the full width of the $J/\psi$. 
However, since the $J/\psi$ couples to conserved lepton and hyperon currents, the
contribution from the $P_\mu P_\nu$ term vanishes. 
In conclusion, the matrix element for $e^+e^-$ annihilation through a photon 
will be structurally identical to that for annihilation through a $J/\psi$ provided 
we make the replacement 
\begin{equation}
	\frac{e_\psi e_g}{s-m_\psi^2+im_\psi^2\Gamma(\psi)}\rightarrow \frac{e_{em}^2}{s},
\end{equation}
where $e_{em}$ is the electric charge.



\section{Cross section for $e^+e^-\rightarrow \Lambda(s_1) \bar{\Lambda}(s_2)$} \label{Sect3}

Our first task is to calculate the cross-section distribution for $e^+e^-$ annihilation 
into polarized hyperons. From the squared matrix element $|{\cal{M}}|^2$ for this process 
we remove  a factor $\cal{K}_\psi$, to get 
\begin{equation}
	\rd \sigma= \frac{1}{2s} \,{\cal{K}_\psi}\, |{\cal{M}}_{red}|^2\,\,
	   \textrm{dLips}(k_1+k_2;p_1,p_2)	 ,
\end{equation}
with $\textrm{dLips}$ the  phase-space factor,  with $s=P^2$, and with
\begin{equation}
{\cal{K}_\psi} =\frac{e_\psi^2e_g^2}{(s-m_\psi^2)^2+m_\psi^2\Gamma^2(m_\psi)}.
\end{equation}
The square of the reduced matrix element can be factorized as
\begin{equation}
	\left| {\cal{M}}_{red}(e^+e^-\rightarrow \Lambda(s_1)\overline{\Lambda}(s_2))\right|^2 =
	 L\cdot K(s_1,s_2) , \label{LLredM}
\end{equation}
with $L(k_1,k_2)$ and $K(p_1,p_2;s_1,s_2)$  lepton and hadron tensors,  and  $s_1$ and $s_2$  
hyperon spin four-vectors. 

Lepton tensor including averages over lepton spins;
\begin{align}
	L_{\nu\mu}(k_1,k_2)=&\,  \fourth {\rm Tr}\big[ \gamma_\nu \, 
	/ \!\!\!k_1 \gamma_\mu\,  / \!\!\!k_2  \big] \nonumber \\
	 =&\, k_{1\nu}k_{2\mu}+ k_{2\nu}k_{1\mu} -\half s g_{\nu\mu}.
\end{align}

Hadron tensor for polarized hyperons;
\begin{align}
 K_{\nu\mu}(s_1,s_2)={\rm Tr}\Big[ &\overline{\Gamma}_\nu^\Lambda(/\!\!\!p_1+M)\half(1+\gamma_5/\!\!\! s_1) 
\nonumber\\
 & \times\Gamma_\mu^\Lambda
(/ \!\!\! p_2-M)\half(1+\gamma_5/\!\!\! s_2)
\Big]/e_g^2 ,
\end{align}
with  $p_1$ and $s_1$ momentum and spin for the Lambda hyperon and $p_2$ and $s_2$ correspondingly 
for the anti-Lambda hyperon. The trace itself is symmetric in the two hyperon variables.

The spin four-vector $s(\mathbf{p},\mathbf{n})$ of a hyperon of mass $M$, three-momentum $\mathbf{p}$, 
and spin direction $\mathbf{n}$ in its rest system, is
\begin{equation}
	s(\mathbf{p},\mathbf{n})=\frac{n_{\parallel}}{M}(|\mathbf{p}|; E \hat{\mathbf{p}})+(0;\mathbf{n}_\bot).
	\end{equation}
	Here, longitudinal and transverse designations refer to the $\hat{\mathbf{p}}$ direction; 
	$n_{\parallel}=\mathbf{n}\cdot\hat{\mathbf{p}}$ and
$\mathbf{n}_\bot=\mathbf{n} -\hat{\mathbf{p}}(\mathbf{n}\cdot \hat{\mathbf{p}})$ are 
parallel and transverse components  of the spin vector $\mathbf{n}$. Also, observe that 
the four-vectors $p$ and $s$ are orthogonal, i.e.~$p\cdot s(p)=0$.
	
For the evaluation of the matrix element we turn to the global c.m.~system where kinematics simplifies. 
Here, three-momenta $\mathbf{p}$ and $\mathbf{k}$ are defined such that
	\begin{align}
	\mathbf{p}_1  &=  - \mathbf{p}_2= \mathbf{p} , \\
	\mathbf{k}_1  &= - \mathbf{k}_2 = \mathbf{k}, 
\end{align}
	and scattering angle by,
	\begin{equation}
		\cos\theta= \hat{\mathbf{p}}\cdot \hat{\mathbf{k}}.
	\end{equation}
	The phase-space factor becomes
\begin{equation}
	 \textrm{dLips}(k_1+k_2;p_1,p_2)	= \frac{p}{32\pi^2 k}\,\rd \Omega, 
\end{equation}
with $p=|\mathbf{p}| $ and $k=|\mathbf{k}| $.

The matrix element in Eq.(\ref{LLredM}) can be written as a sum of four terms that depend 
on the hyperon spin directions
in their respective rest systems,  $\mathbf{n}_1$ and $\mathbf{n}_2$, 
\begin{align}
	\left| {\cal{M}}_{red}(e^+e^-\right. &\rightarrow \left.\Lambda(s_1)\overline{\Lambda}(s_2))\right|^2= \, sD(s)\, \Big[
	H^{00}(0,0)+H^{05}(\mathbf{n}_1,0) \nonumber\\
	&\quad +H^{50}(0,\mathbf{n}_2)+ H^{55}(\mathbf{n}_1,\mathbf{n}_2)\Big].\label{Hintro}
\end{align}
The polarization distributions $H^{ab}$ are each expressed in terms  of structure functions 
that depend on the scattering angle $\theta$, the ratio function $\alpha(s)$, and
the phase function $\Delta\Phi(s)$. There are six such structure functions,
\begin{eqnarray}
{\cal{R}} &=& 1 +\alpha  \cos^2\!\theta, \label{DefR}\\
  {\cal{S}} &=& \sqrt{1-\alpha^2}\sin\theta\cos\theta\sin(\Delta\Phi), \label{DefS}\\
	{\cal{T}}_1 &=& \alpha + \cos^2\!\theta, \\
	{\cal{T}}_2 &=& -\alpha\sin^2\!\theta,  \\
	{\cal{T}}_3 &=& 1+\alpha, \\
	{\cal{T}}_4 &=& \sqrt{1-\alpha^2}\cos\theta\cos(\Delta\Phi). \label{RSTdef}
\end{eqnarray}
The definitions and notations are slightly different from those of Ref.~\cite{GF2}. 
In particular, a factor $sD(s)$ has been pulled out from the structure functions, 
and placed in front of the sum of the polarization distributions of Eq.(\ref{Hintro}).

The polarization distributions $H^{ab}$ are, 
\begin{eqnarray}
	H^{00} &=&   {\cal{R}}  \label{H00def}\\
	H^{05} &=&  {\cal{S}} 	   \left[ \frac{1}{\sin\theta} ( \hat{\mathbf{p}}\times 
	  \hat{\mathbf{k}})\cdot \mathbf{n}_{1}
	      \right]  \label{Lpol}\\      
	H^{50} &=& {\cal{S}} 
	   \left[ \frac{1}{\sin\theta} ( \hat{\mathbf{p}}\times \hat{\mathbf{k}})\cdot \mathbf{n}_{2} 
	             \right] \label{ALpol}\\
	H^{55} &=& \bigg\{  {\cal{T}}_1 
	  \mathbf{n}_{1} \cdot \hat{\mathbf{p}}\mathbf{n}_{2} \cdot \hat{\mathbf{p}}
	   + {\cal{T}}_2  \mathbf{n}_{1\bot} \cdot \mathbf{n}_{2\bot} 
		\nonumber\\
		 &&  \qquad + {\cal{T}}_3 \mathbf{n}_{1\bot} \cdot \hat{\mathbf{k}}\mathbf{n}_{2\bot}
		\cdot \hat{\mathbf{k}} \nonumber\\
			&&  \qquad + {\cal{T}}_4  \bigg( \mathbf{n}_{1} \cdot \hat{\mathbf{p}}\mathbf{n}_{2\bot} \cdot \hat{\mathbf{k}} 
			 + \mathbf{n}_{2} \cdot \hat{\mathbf{p}}\mathbf{n}_{1\bot} \cdot \hat{\mathbf{k}} 
			  \bigg)  \bigg\} \label{H55def}
\end{eqnarray}
Transverse components $\mathbf{n}_{1\bot}$ and $\mathbf{n}_{2\bot}$ are orthogonal 
to the Lambda hyperon momentum $\mathbf{p}$ in the global c.m.~system. Also, transverse 
$\mathbf{n}_{\bot}$ and longitudinal $n_\|=\hat{\mathbf{p}}\cdot\mathbf{n}$ 
polarization components enter differently, since they transform 
differently under Lorentz transformations.

All polarization observables, single and double, can be directly read off Eqs.(\ref{H00def}-\ref{H55def}), 
and there are no other possibilities. The set of scalar products involving 
$\mathbf{n}_{1}$ and $\mathbf{n}_{2}$ is complete. As an example, the Lambda-hyperon polarization 
is obtained from Eq.(\ref{Lpol}) which shows that the polarization is
directed along the normal to the scattering plane, $ \hat{\mathbf{p}}\times 
	  \hat{\mathbf{k}}$, and that the value of the polarization is
\begin{equation}
	P_\Lambda(\theta)=\frac{{\cal{S}}}{{\cal{R}}} =
\frac{\sqrt{1-{\alpha}^2}\cos\theta\sin\theta}{1+\alpha\cos^2\!\theta}\sin({\Delta\Phi})\label{Polbelopp}
\end{equation}
From Eq.(\ref{ALpol}) we conclude that the polarization of the anti-Lambda is
exactly the same, but then one should remember that $\mathbf{p}$ is the momentum of the Lambda hyperon 
but $-\mathbf{p}$  that of the anti-Lambda.

%


\section{Folding of distributions} 

Our next task is to calculate the cross-section distribution for $e^+e^-$ annihilation 
into hyperon pairs, followed by the hyperon decays into nucleon-pion pairs. 
This reaction is described by the connected diagram of Fig.~2.

Again, we extract a prefactor, ${\cal{K}}={\cal{K}}_\psi{\cal{K}}_1{\cal{K}}_2$, from 
the squared matrix element, writing
\begin{equation}
	|{\cal{M}}|^2=   {\cal{K}}     |{\cal{M}}_{red}|^2 .
\end{equation}
The prefactor originates,  as before with the propagator denominators.
Due to the smallness of the hyperon widths each of the hyperon propagators 
can, after squaring, be approximated as,
\begin{equation}
	{\cal{K}}_i =
	\frac{1}{(p_i^2-M^2)^2+M^2\Gamma^2(M)}  = 
	  \frac{2\pi}{2M\Gamma(M)} \, \delta(p_i^2-M^2) .
\end{equation}
Effectively, this approximation puts the hyperons on their mass shells. 

	\begin{figure}[ht]
\begin{center}
\resizebox{0.38\textwidth}{!}
{ \includegraphics{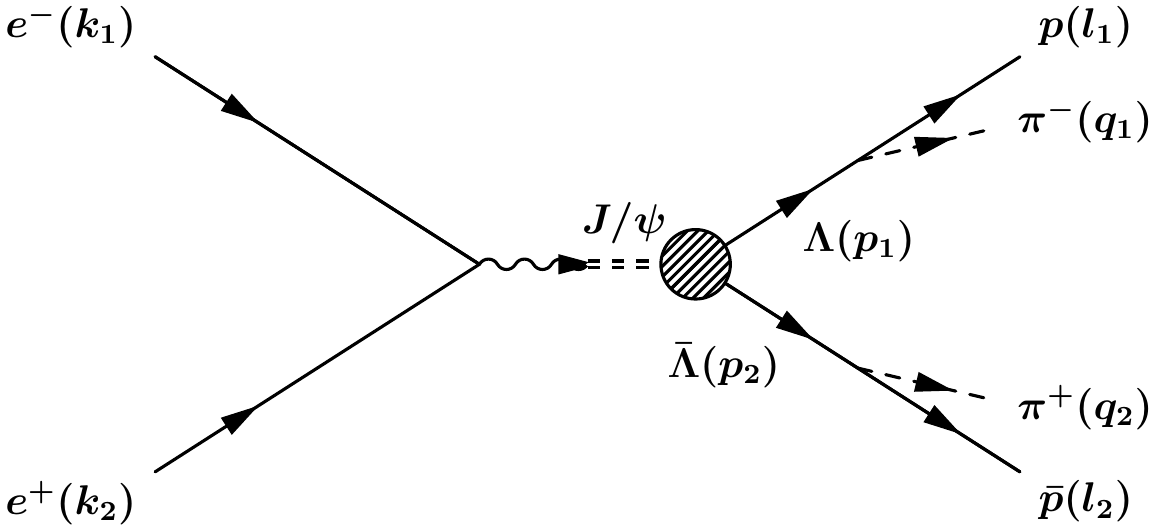}     }
\caption{Graph describing the reaction 
$e^+ e^- \rightarrow  \Lambda(\rightarrow p\pi^-)\bar{\Lambda}(\rightarrow\bar{p}\pi^+)$.}
\label{F2-fig}
\end{center}
\end{figure}

	Hyperon-decay distributions are obtained by a folding calculation, whereby  hyperon-production
	and -decay distributions are multiplied together and
	averaged over the intermediate hyperon-spin directions. It was proved in Ref.\cite{GF1}
	that the folding prescription gives the 
	same result as the evaluation of the connected-Feynman-diagram expression.
 Hence, summing over final hadron spins,  
	\begin{align}
& |{\cal{M}}|^2 =\sum_{\pm s_1,\pm s_2}
\left\langle 	\left| {\cal{M}}(e^+ e^- \rightarrow \Lambda(s_1) \bar{\Lambda}(s_2))\right|^2 \right.
 \nonumber \\
& \qquad  \left.
\times
 \left| {\cal{M}}(\Lambda(s_1) \rightarrow  p\pi^-)\right|^2
\left| {\cal{M}}(\bar{\Lambda}(s_2) \rightarrow \bar{p}\pi^+)\right|^2  
\right\rangle_{\mathbf{n}_1\mathbf{n}_2} .\label{Fold-def}
\end{align}
Production and decay distributions are,
\begin{align}
&	\left| {\cal{M}}(e^+ e^- \rightarrow \Lambda(s_1) \bar{\Lambda}(s_2))\right|^2=L\cdot       K(s_1,s_2)  ,\\
& \left| {\cal{M}}(\Lambda(s_1) \rightarrow  p\pi^-)\right|^2 =
  R_\Lambda\left[ 1 -
	   \alpha_1  l_1\cdot s_1/l_\Lambda \right]  ,\label{Ldecay}\\
& \left| {\cal{M}}(\bar{\Lambda}(s_2) \rightarrow  \bar{p}\pi^+)\right|^2 =
  R_\Lambda\left[ 1 -
	   \alpha_2   l_2\cdot s_2/l_\Lambda \right],\label{Lbardecay}
\end{align}
with $l_\Lambda$ the decay momentum in the Lambda rest system. 
$R_\Lambda$	is determined by the Lambda decay rate.

The notation in Eq.(\ref{Ldecay}) is the following; $s_1$ denotes the Lambda four-spin vector,
$l_1$  the four-momentum of the decay proton, and $\alpha_1$ the decay-asymmetry parameter. 
Similarly for the anti-Lambda hyperon parameters of Eq.(\ref{Lbardecay}).

We evaluate the hyperon-decay distributions in the hyperon-rest systems, where
\begin{align}
& \left| {\cal{M}}(\Lambda(s_1) \rightarrow  p\pi^-)\right|^2 =
  R_\Lambda\left[ 1 +
	   \alpha_1  \hat{ \mathbf{l}}_1\cdot \mathbf{n}_1\right]  ,\label{LIIdecay}\\
& \left| {\cal{M}}(\bar{\Lambda}(s_2) \rightarrow  \bar{p}\pi^+)\right|^2 =
  R_\Lambda\left[ 1+
	   \alpha_2  \hat{ \mathbf{l}}_2\cdot \mathbf{n}_2 \right],\label{LIIbardecay}
\end{align}
where $\hat{ \mathbf{l}}_1={\mathbf{l}}_1/l_\Lambda$ is the unit vector in the direction  
of the proton momentum in the Lambda-rest system, and correspondingly for the anti-Lambda case.

Angular averages in Eq.(\ref{Fold-def}) are calculated according to the prescription
\begin{equation}
\big{\langle } ( \mathbf{n}\cdot \mathbf{l} ) \mathbf{n}\big{\rangle }_{\mathbf{n}}   =\mathbf{l}.
 \end{equation} 

The folding of the production distributions, Eqs.(\ref{H00def}-\ref{H55def}), with the 
decay distributions, Eqs.(\ref{LIIdecay}-\ref{LIIbardecay}), yields 
\begin{equation}
	|{\cal{M}}_{red}|^2=sD(s) R_{\Lambda}^2 \bigg[ G^{00} +G^{05} +G^{50} + G^{55}\bigg] ,\label{MexpG}
\end{equation}
with the  $G^{ab}$ functions defined as
\begin{eqnarray}
	G^{00} &=&   {\cal{R}}, \label{G00}\\
	G^{05} &=&\alpha_1  {\cal{S}}
	   \left[ \frac{1}{\sin\theta} ( \hat{\mathbf{p}}\times \hat{\mathbf{k}})\cdot \hat{\mathbf{l}}_{1}
	      \right] , \label{G05}\\      
	G^{50} &=& \alpha_2 {\cal{S}} 
	   \left[ \frac{1}{\sin\theta} ( \hat{\mathbf{p}}\times \hat{\mathbf{k}})\cdot \hat{\mathbf{l}}_{2} 
	             \right], \label{G50}\\
	G^{55} &=& \alpha_1\alpha_2 \bigg\{  {\cal{T}}_1 
	  \hat{\mathbf{l}}_{1} \cdot \hat{\mathbf{p}}\hat{\mathbf{l}}_{2} \cdot \hat{\mathbf{p}}
	   + {\cal{T}}_2  \hat{\mathbf{l}}_{1\bot} \cdot \hat{\mathbf{l}}_{2\bot} \nonumber \\
		 && \qquad \quad + {\cal{T}}_3 \hat{\mathbf{l}}_{1\bot} \cdot \hat{\mathbf{k}}\hat{\mathbf{l}}_{2\bot}
		\cdot \hat{\mathbf{k}} \nonumber
		 \\
			&& \qquad \quad + {\cal{T}}_4  \bigg( \hat{\mathbf{l}}_{1} \cdot \hat{\mathbf{p}}\hat{\mathbf{l}}_{2\bot} \cdot \hat{\mathbf{k}} 
			 + \hat{\mathbf{l}}_{2} \cdot \hat{\mathbf{p}}\hat{\mathbf{l}}_{1\bot} \cdot \hat{\mathbf{k}} 
			  \bigg)  \bigg\}.\label{G55}
\end{eqnarray}
Thus, we conclude the connection between joint-hadron production and joint-hadron decay distributions simply to be,
\begin{equation}
	G^{ab}(\hat{\mathbf{l}}_{1}; \hat{\mathbf{l}}_{2})= 
	H^{ab}(\mathbf{n}_{1}\rightarrow \alpha_1 \hat{\mathbf{l}}_{1}; 
	        \mathbf{n}_{2}\rightarrow \alpha_2  \hat{\mathbf{l}}_{2}).
\end{equation}

We repeat the notation; $\mathbf{p}$ and $\mathbf{k}$ are momenta of Lambda and electron in the 
  global c.m.\ system;
 $\mathbf{l}_{1}$ and $\mathbf{l}_{2}$ are  momenta of proton and anti-proton in Lambda and 
anti-Lambda rest systems; orthogonal means orthogonal to  $\mathbf{p}$; and 
structure functions ${\cal{R}}$, ${\cal{S}}$, and ${\cal{T}}$ are functions 
of $\theta$, $\alpha$, and $\Delta\Phi$. The angular functions multiplying the structure functions form 
a set of seven mutually orthogonal functions, when integrated over the proton
and anti-proton decay angles.
 
%
%
%

%

\section{Cross section for $e^+e^-\rightarrow \Lambda(\rightarrow 
                 p\pi^-) \bar{\Lambda}(\rightarrow \bar{p}\pi^+)$} 
  
	Our last task is to find the properly normalized cross-section distribution. 
	We start from the general expression,
\begin{equation}
	\rd \sigma= \frac{1}{2s} \,{\cal{K}}\, |{\cal{M}}_{red}|^2\,\,
	   \textrm{dLips}(k_1+k_2;q_1,l_1,q_2,l_2)	 ,	 
\end{equation}
with $ \textrm{dLips}$ the phase-space density for four final-state particles.
The prefactor ${\cal{K}}$ contains on the mass shell delta functions for
the two hyperons. This effectively separates the phase space into 
production and decay parts. Repeating the manipulations of Ref.\cite{GF1} 
we get  
\begin{eqnarray}
\rd \sigma  & = & \frac{1}{64\pi^2 }\frac{p}{k}
	  \frac{\alpha_{g}\alpha_\psi}{(s-m_\psi^2)^2+m_\psi^2\Gamma^2(\psi)}
		\frac{\Gamma_{\Lambda}\Gamma_{\bar{\Lambda}}}{\Gamma^2(M)}\cdot
	   \nonumber \\	
	&& \cdot \left(D(s)\sum_{a,b} G^{ab} \right)\, \rd \Omega_{\Lambda}
	   \rd \Omega_1\rd \Omega_2 , \label{Sigmaster}
\end{eqnarray}
with $k$ and $p$	 the initial- and final-state momenta; 
$\Omega_{\Lambda}$ the hyperon scattering angle in the global c.m.\ system;  
$\Omega_{1}$ and $\Omega_{2}$ decay angles measured in the rest systems of $\Lambda$ 
and $\bar{\Lambda}$; 
$\Gamma_{\Lambda}$ and $\Gamma_{\bar{\Lambda}}$  channel widths;
and $\Gamma(M)$ and $\Gamma(\psi)$  total widths.

Integration over the  angles $\Omega_1$  makes the contributions from 
the functions $G^{05}$ and $G^{55}$ disappear  \cite{GF1}, and correspondingly for 
the angles $\Omega_{2}$. Integration over both angular variables results in
the cross-section distribution for the reaction 
$e^+ e^- \rightarrow J/\psi \rightarrow \Lambda\bar{\Lambda}$.

Suppose we integrate over the angles $\Omega_{2}$. Then, the predicted hyperon-decay distribution becomes 
 proportional to the sum 
\begin{align}
	G^{00} + G^{05} &= {\cal{R}} \left( 1 + \alpha_1\mathbf{P}_\Lambda \cdot \hat{\mathbf{l}}_{1} \right),\\
	 P_\Lambda  &= \frac{{\cal{S}}}{{\cal{R}}}, \label{pol2}
\end{align}
with the polarization $P_{\Lambda}$ as in Eq.(\ref{Polbelopp}), 
and the polarisation vector $ \mathbf{P}_\Lambda$  directed along the normal to the scattering plane 

%
%

\section{Differential distributions}

We first define our coordinate system.
The scattering plane with the vectors $\mathbf{p}$ and $\mathbf{k}$ 
	make up the $xz$-plane, with the $y$-axis  along the normal to the scattering plane. 
	We choose a right-handed coordinate system with basis vectors 
	\begin{eqnarray}
	\mathbf{e}_z  &=&  \hat{\mathbf{p}}, \label{zunity}\\
	\mathbf{e}_y  &=& \frac{1}{\sin\theta } ( \hat{\mathbf{p}}\times \hat{\mathbf{k}} ) ,\label{yunity} \\
	\mathbf{e}_x  &=& \frac{1}{\sin\theta } ( \hat{\mathbf{p}}\times \hat{\mathbf{k}} ) 
	 \times\hat{\mathbf{p}}.\label{xunity}
\end{eqnarray}
Expressed in terms of them the initial-state momentum 
\begin{equation}
	\hat{\mathbf{k}}= \sin\theta\,  \mathbf{e}_x +\cos\theta\,	\mathbf{e}_z  .
\end{equation}

This coordinate system is used for fixing the directional angles of 
the decay proton in the Lambda rest system, and the decay anti-proton in the anti-Lambda rest system.
 The spherical angles for the proton are 
$\theta_1$ and $\phi_1$, and the components of the unit vector in 
direction of the decay-proton momentum are,
\begin{equation}
	\hat{\mathbf{l}}_1=(\cos \phi_1 \sin \theta_1,  \sin \phi_1 \sin \theta_1, \cos \theta_1),
\end{equation}
so that
\begin{equation}
	\hat{\mathbf{l}}_{1\bot}=(\cos \phi_1 \sin \theta_1,  \sin \phi_1 \sin \theta_1, 0).
\end{equation}
The momentum of the decay proton is by definition  $\mathbf{l}_1=l_\Lambda \hat{\mathbf{l}}_1$.
This same coordinate system is used for defining the directional angles of 
the decay anti-proton in the anti-Lambda rest system, with spherical angles 
$\theta_2$ and $\phi_2$.

Now, we have all ingredients  needed for the calculation of the $G$ functions of Eqs.(\ref{G00}-\ref{G55}), 
the functions that in the end determine the cross-section distributions.

An event of the reaction 
$e^+e^-\rightarrow \Lambda(\rightarrow p\pi^-) \bar{\Lambda}(\rightarrow \bar{p}\pi^+)$
 is specified by the five dimensional vector 
${\boldsymbol{\xi}}=(\theta,\Omega_{1},\Omega_{2})$, and the differential-cross-section distribution as
summarized  by Eq.(\ref{MexpG}) reads,
\[
{\rd\sigma}\propto {\cal{W}}({\boldsymbol{\xi}})\ {\rd\!\cos\theta\ \rd\Omega_{1}\rd\Omega_{2}}.
\]
At the moment, we are not interested in the absolute normalization of the differential distribution. 
The differential-distribution function ${\cal{W}}({\boldsymbol{\xi}})$ 
is obtained from Eqs.(\ref{G00}-\ref{G55}) and can be expressed as,
\begin{equation}
\begin{split}
{\cal{W}}({\boldsymbol{\xi}})=&{\cal{F}}_0({\boldsymbol{\xi}})+{{{\alpha}}}{\cal{F}}_5({\boldsymbol{\xi}})\\
+&{{\alpha_1\alpha_2}}\left({\cal{F}}_1({\boldsymbol{\xi}})
+\sqrt{1-{{\alpha}}^2}\cos({{\Delta\Phi}}){\cal{F}}_2({\boldsymbol{\xi}})
+{{\alpha}}{\cal{F}}_6({\boldsymbol{\xi}})\right)\\
+&\sqrt{1-{{\alpha}}^2}\sin({{\Delta\Phi}})
\left({{\alpha_1}}{\cal{F}}_3({\boldsymbol{\xi}})
+{{\alpha_{2}}}{\cal{F}}_4({\boldsymbol{\xi}})\right),  \label{eqn:pdf}
\end{split}
\end{equation}
using a set of seven angular 
functions ${\cal{F}}_k({\boldsymbol{\xi}})$ defined as:
\begin{align}
	{\cal{F}}_0({\boldsymbol{\xi}}) =&1 \nonumber\\
	{\cal{F}}_1({\boldsymbol{\xi}}) =&{\sin^2\!\theta}\sin\theta_1\sin\theta_2\cos\phi_1\cos\phi_2+
{\cos^2\!\theta}\cos\theta_1\cos\theta_2\nonumber\\
	{\cal{F}}_2({\boldsymbol{\xi}}) =&{\sin\theta\cos\theta}\left(\sin\theta_1\cos\theta_2\cos\phi_1+
\cos\theta_1\sin\theta_2\cos\phi_2\right)\nonumber\\
	{\cal{F}}_3({\boldsymbol{\xi}}) =&{\sin\theta\cos\theta}\sin\theta_1\sin\phi_1\nonumber\\
	{\cal{F}}_4({\boldsymbol{\xi}}) =&{\sin\theta\cos\theta}\sin\theta_2\sin\phi_2\nonumber\\
	{\cal{F}}_5({\boldsymbol{\xi}}) =&{\cos^2\!\theta}\nonumber\\
	{\cal{F}}_6({\boldsymbol{\xi}}) =&\cos\theta_1\cos\theta_2 -\sin^2\!\theta\sin\theta_1\sin\theta_2\sin\phi_1\sin\phi_2.
\end{align}

The differential distribution of Eq.~(\ref{eqn:pdf}) involves two parameters related to the $e^+e^-\to\Lambda\bar{\Lambda}$ process that can be determined by data: the ratio of form factors $\alpha$,  and the relative phase of form factors $\Delta\Phi$. In addition, the distribution function ${\cal{W}}({\boldsymbol{\xi}})$ can be used to extract
separately $\Lambda$ and ${\bar{\Lambda}}$ decay-asymmetry parameters:  $\alpha_1$ and  $\alpha_{2}$, and 
hence allowing  a direct test of CP conservation in the hyperon decays.

The  term proportional to $\sin({{\Delta\Phi}})$ in Eq.~(\ref{eqn:pdf})  originates with  
 Eqs.(\ref{G05}) and (\ref{G50}), and  can be rewritten as,
\[
{\cal{S}}(\theta)\left(\alpha_1\sin\theta_1\sin\phi_1+
\alpha_2\sin\theta_2\sin\phi_2\right) ,
\]
with the structure function ${\cal{S}}$ defined by Eq.(\ref{DefS}). The relation between the structure functions 
and the polarization $P_\Lambda(\theta)$ was discussed in Sect.~\ref{Sect3}, where it was shown that 
the polarization, $P_\Lambda(\theta)$ of Eq.(\ref{Polbelopp}), and the polarization vector, $\mathbf{e}_y$, are the same
 for Lambda and anti-Lambda hyperons. This information tells us that $\Lambda$ is polarized 
along the normal  to the production
plane, and that the polarization vanishes at $\theta=0^\circ$, $90^\circ$ and
$180^\circ$. The maximum value of the polarization is for
$\cos\theta=\pm 1/(2+\alpha)$, and
$|P_\Lambda(\theta)|<\frac{2}{3}\sin({\Delta\Phi})$.

It should be
stressed that the simplified distributions used in previous analyses, 
such as Ref.\cite{BESa},
assume the hyperons to be unpolarized and therefore terms
containing $P_\Lambda(\theta)$ are missing. In fact, such decay
distributions,  only permit the  determination of two parameters:
the ratio of form factors $\alpha$, and the product of hyperon-asymmetry parameters $\alpha_1\alpha_2$. 

In our opinion, the
formulas presented in this letter should be employed for the exclusive
analysis of the new BESIII data \cite{Ablikim17a}. Due to huge and clean
data samples: $(440675\pm670)$ $J/\psi\to\Lambda{\bar{\Lambda}}$ and
$(31119\pm 187)$ $\psi(3686)\to\Lambda{\bar{\Lambda}}$, precision values 
for the decay-hadronic-form factors could be extracted as well as precision 
values for $\Lambda$ and $\bar{\Lambda}$ decay-asymmetry parameters. The formulas presented
could easily be generalized to neutron decays of the $\Lambda$ and to production of
other $J=1/2$ hyperons with analogous decay modes.

\section*{Appendix}

The coupling of the
initial-state leptons to the $J/\psi$ vector meson is  determined by the decay $J/\psi\rightarrow e^+e^-$. 
Assuming  the  decay to go via an intermediate photon, Fig.1b, we can safely ignore any tensor
coupling. The vector coupling of the $J/\psi$ to leptons is therefore the 
same as for the photon, if replacing the electric charge $e_{em}$ by a coupling 
strength $e_\psi$. From the decay $J/\psi\rightarrow e^+e^-$ one derives
\begin{equation}
	\alpha_{\psi}=e_\psi^2/4\pi=3\Gamma(J/\psi\rightarrow e^+e^-)/m_\psi. \label{epsi}
\end{equation}

 In a similar fashion  we relate the strength $e_g$ of $J/\psi$ coupling to the hyperons 
to the  decay $J/\psi\rightarrow \Lambda\bar{\Lambda}$. 
In analogy with Eq.(\ref{epsi})
we get
\begin{align}
	\alpha_{g}=e_{g}^2/4\pi=& 3 
	  \left(  (1+{2M^2}/{m_\psi^2})\sqrt{1-{4M^2}/{m_\psi^2} }\,   \right)^{-1}\nonumber\\
	 &\times \Gamma(J/\psi\rightarrow \Lambda\bar{\Lambda})/m_\psi .
\end{align}
When the $\Lambda$ mass $M$ is replaced by the lepton mass $m_l=0$ we recover Eq.(\ref{epsi}).
%
%

\section*{Acknowledgments}
We are grateful to Tord Johansson who provided the inspiration for this work.



\end{document}